\documentclass[acmsmall, natbib=false, screen, nonacm]{acmart}

\authorsaddresses{} %

\usepackage[utf8]{inputenc} %
\usepackage{verbatim}
\usepackage{bm}
\usepackage[english]{babel}
\usepackage{pslatex}
\usepackage{graphicx}
\usepackage{wrapfig}
\usepackage{framed}
\usepackage{booktabs} %
\usepackage{tabularx}
\usepackage{multicol}
\usepackage{multirow}
\usepackage[textsize=small,shadow]{todonotes}  %
\usepackage{color}
\usepackage{xcolor}
\usepackage{colortbl}
\usepackage[normalem]{ulem} %
\usepackage{xspace} 
\usepackage{hyperref}
\usepackage{markdown}

\makeatletter
\DeclareRobustCommand\onedot{\futurelet\@let@token\@onedot}
\def\@onedot{\ifx\@let@token.\else.\null\fi\xspace}

\def\eg{\emph{e.g}\onedot} 
\def\ie{\emph{i.e}\onedot}

\newif\ifdraft
\draftfalse
\ifdraft
	\def\an#1{{\color{MidnightBlue}[AN] #1}}
	\def\me#1{{\color{RedOrange}[ME] #1}}
        \def\cc#1{{\color{PineGreen}[EP] #1}}

\else
	\def\an#1{}
	\def\me#1{}
        \def\cc#1{}
        \def\jv#1{}
        \def\eg#1{}
\fi

\AtBeginDocument{%
  }

\setcopyright{cc}
\setcctype{by-sa}
\copyrightyear{2025}
\acmYear{2025}
\acmDOI{XXXXXXX.XXXXXXX}

\RequirePackage[
  datamodel=acmdatamodel,
  style=acmauthoryear,
  ]{biblatex}

\begin{document}

\title{Generative AI and the Future of the Digital Commons}
\subtitle{Five Open Questions and Knowledge Gaps}
\titlenote{\textbf{Disclaimer:} The views expressed in this paper are those of the authors and may not, under any circumstances, be regarded as an official position of the European Commission.}

\author{Arman Noroozian}
\affiliation{%
  \institution{European Commission, Joint Research Centre (JRC)}
  \city{Brussels}
  \country{Belgium}
}
\authornote{\textbf{Note:} Except for the main author, all other contributing authors have been included in alphabetical order}

\author{Lorena Aldana}
\affiliation{%
  \institution{Europeana Foundation}
  \city{The Hague}
  \country{Netherlands}
}
\author{Marta Arisi}
\affiliation{%
  \institution{Sciences Po Law School}
  \city{Paris}
  \country{France}
}
\author{Hadi Asghari}
\affiliation{%
  \institution{Technische Universität Berlin}
  \city{Berlin}  
  \country{Germany}
}
\author{Renata Avila}
\affiliation{%
  \institution{Open Knowledge Foundation}
  \city{London}
  \country{United Kingdom}
}
\author{Pietro Giovanni Bizzaro}
\affiliation{%
  \institution{Fondazione Bruno Kessler and Università di Trento}
  \city{Trento}
  \country{Italy}
}
\author{Ramya Chandrasekhar}
\affiliation{%
  \institution{Centre for Internet and Society, CNRS}
  \city{Paris}
  \country{France}
}
\author{Cristian Consonni}
\affiliation{%
  \institution{European Commission, Joint Research Centre (JRC)}
  \city{Ispra}
  \country{Italy}
}
\author{Deborah	De Angelis}
\affiliation{%
  \institution{Creative Commons Italy Chapter}
  \city{Florence}
  \country{Italy}
}
\author{Francesca De Chiara}
\affiliation{%
  \institution{CNR ISMN}
  \city{Bologna}
  \country{Italy}
}
\author{Maria del Rio-Chanona}
\affiliation{%
  \institution{Computer Science Dept. and Center for Artificial Intelligence, UCL}
  \city{London}
  \country{United Kingdom}
}
\author{Melanie Dulong de Rosnay}
\affiliation{%
  \institution{Center for Internet and Society, CNRS}
  \city{Paris}
  \country{France}
}
\author{Maria Eriksson}
\affiliation{%
  \institution{European Commission, Joint Research Centre (JRC)}
  \city{Seville}
  \country{Spain}
}
\author{Frederic Font}
\affiliation{%
  \institution{Universitat Pompeu Fabra}
  \city{Barcelona}
  \country{Spain}
}
\author{Emilia Gomez}
\affiliation{%
  \institution{European Commission, Joint Research Centre (JRC)}
  \city{Seville}
  \country{Spain}
}
\author{Valérian Guillier}
\affiliation{%
  \institution{Centre for Internet and Society, CNRS}
  \city{Paris}
  \country{France}
}
\author{Lisa Gutermuth}
\affiliation{%
  \institution{Mozilla Foundation}
  \city{Berlin}
  \country{Germany}
}
\author{David Hartmann}
\affiliation{%
  \institution{Weizenbaum Institute}
  \city{Berlin}
  \country{Germany}
}
\author{Lucie-Aimée Kaffee}
\affiliation{%
  \institution{Hugging Face}
  \city{Berlin}
  \country{Germany}
}
\author{Paul Keller}
\affiliation{%
  \institution{Open Future Foundation and COMMUNIA Association}
  \city{Amsterdam}
  \country{The Netherlands}
}
\author{Felix Stalder}
\affiliation{%
  \institution{Zurich University of the Arts}
  \city{Zurich}
  \country{Switzerland}
}
\author{Jo\~{a}o Vinagre}
\affiliation{%
  \institution{European Commission, Joint Research Centre (JRC)}
  \city{Seville}
  \country{Spain}
}
\author{Denny Vrandečić}
\affiliation{%
  \institution{Wikimedia Foundation}
  \city{San Francisco}
  \country{USA}
}
\author{Amanda Wasielewski}
\affiliation{%
  \institution{Department of ALM, Uppsala University}
  \city{Uppsala}
  \country{Sweden}
}

\renewcommand{\shortauthors}{Noroozian et al.}



\received{19 June 2025}
\received[revised]{XX XXXX XXXX}
\received[accepted]{XX XXXX XXXX}

\maketitle

\section*{}

On April 1st 2025, the Wikimedia Foundation -- which hosts the largest and most widely used archives of open knowledge in the world -- reported in a blog post\footnote{\href{https://diff.wikimedia.org/2025/04/01/how-crawlers-impact-the-operations-of-the-wikimedia-projects/}{How crawlers impact the operations of the Wikimedia projects}, retrieved May 9th, 2025.} that web crawlers are impacting their operations to the point that they are experiencing difficulties to cater to their human users during traffic spikes. Following the rise of AI, 65 percent of the Wikimedia project's most expensive traffic now originates from web crawlers attempting to bulk read their content, and since January 2024, the organization has witnessed a bandwidth growth of 50 percent for downloads of multimedia content from their open catalogs. ``Our content is free,'' Wikimedia stated, ``but our infrastructure is not,'' raising concerns about the increased resources needed to handle high-intensity traffic originating from AI crawlers. 

Wikimedia's worries about serving a data-intensive AI industry are not unique, but shared with many other organizations, institutions, and groups hosting free and open online content. However, such organizations are often overlooked and struggle to participate in discussions concerning the ethics and socio-technical impacts of AI, which for example tends to focus on the interests of copyright holders, businesses, or everyday online users. While such interests are unquestionably important, the particular challenges that open knowledge and information repositories are facing in the wake of AI also deserve attention. 
Their absence in the public debate is even more striking given that today's generative AI models for images, text, sound, code and video (which we collectively refer to as ``\textit{generative AI}'' or ``\textit{GenAI}''), would not exist without having been trained on content taken from the digital commons: the ensemble of billions of open discussion forums, encyclopedia pages, digitized books and newspapers, repositories of open source code, and open access scientific publications, open digital images, videos, and sounds that are either in the public domain or available under a free and open license. The digital commons not only consist of the digital collections themselves, but also of the communities -- museums, archives, cultural institutions, researchers, and volunteers across the globe --  that contribute with their labor to curate, annotate, catalog and update the collections.
As AI developers train their models, often imparting them with a biased set of values, exacerbated by the compositions of their training content, they also feed on the digital commons which uphold their own norms and values that are not necessarily aligned or upheld when their content gets repurposed and brought out of context.
Currently, content in the digital commons is likely used to train AI models designed for use in fields ranging from medicine, education, and finance, to law enforcement and even military combat.
It is therefore justified to question the nature of this relationship, specifically whether it is symbiotic, and consider \emph{\textbf{how AI development is impacting the digital commons?}}
Does the repurposing of content from the digital commons to train AI models respect and preserve their values? 
What risks, harms, and benefits could GenAI bring to free and open knowledge repositories, the infrastructures behind them, and the communities that help share, care for, and contribute to a thriving digital commons? %
Who benefits from current uses of the digital commons in AI development? At what costs?

As an interdisciplinary group of academics, policy-makers, advocates, and practitioners - having come together for an organized workshop on 
"Generative AI and the Future of the Digital Commons" -  we emphasize the need to recognize that GenAI is likely to have substantial effects on shared and open knowledge repositories in the future. We recognize that GenAI has a positive potential to enrich, strengthen, and contribute to the digital commons, for instance by improving metadata in vast digital collections, translating knowledge repositories and making them accessible to wider audiences, and supporting creative practices whose outputs may later feed back into and make the commons grow. Yet, we also recognize that GenAI is raising a series of unprecedented concerns that are not adequately addressed by existing industry practices and are only just starting to be addressed in governance frameworks, such as the EU AI Act with various gaps to be filled. In what follows, we present five questions that we believe should be placed on the agenda of all researchers, professionals in GLAMs (Galleries, Libraries, Archives, and Museums), non-governmental organizations, policy makers, and AI developers -- including those outside of the EU and active in the ACM community -- who engage with AI and free, open, and shared online data. 

\subsection*{1. How can we ensure that the digital commons are not threatened by undersupply, as people's information-finding needs are increasingly met by closed chatbot services?} 

Recent reports have shown that the introduction of GenAI has lowered incentives to contribute to shared knowledge repositories like open Q\&A forums~\cite{burtch_consequences_2024}. For instance, researchers found that six months after, and due to the release of ChatGPT, activity on Stack Overflow had decreased by 25 percent, relative to counterparts in two markets where ChatGPT is not officially nor readily available due to blocking or technology export restrictions (Russia and China)~\cite{del_rio-chanona_large_2024}. Recent reports estimate that the active weekly users of GenAI chatbots provided by leading industry players may be as high as 400 million~\footnote{\href{https://techcrunch.com/2025/03/06/chatgpt-doubled-its-weekly-active-users-in-under-6-months-thanks-to-new-releases/}{ChatGPT doubled its weekly active users in under 6 months, thanks to new releases}, retrieved on June 5th 2025}, and if these figures persist or continue to grow, it means non-chatbot online activity will continue to decrease in the future. Combined with worries that the integration of GenAI summarization tools in search engines may drastically reduce traffic to third-party websites,
there is a risk that the digital commons may shortly suffer from a lack of supply of new and updated content since less traffic means fewer visitors and likely also fewer content contributors. Importantly, this should not just be an area of concern for supporters of the commons, but also AI developers who are heavily relying on the commons as a source for relevant and diverse data. 

\subsection*{2. How do we address the risk that GenAI -- and its associated extractive data collection practices -- may contribute to a closure of the open web and result in a privatization of otherwise open and shared knowledge?}

In the wake of current AI technology shifts, a large body of evidence points to a growing dissatisfaction with current data collection for AI training purposes. At the moment, the most common ways of signaling dissatisfaction with data crawling and scraping involves specifying restrictions via the \textit{Robots Exclusion Protocol}\footnote{\href{https://datatracker.ietf.org/doc/html/rfc9309}{IETF RFC 9309 - Robots Exclusion Protocol}, retrieved May 9th, 2025.} (\texttt{robots.txt}
files), restrictive \textit{Terms of Service Agreements} (ToS), or employing adversarial techniques to render certain types of content unusable for AI training \footnote{\href{https://glaze.cs.uchicago.edu/}{Glaze and Nightshade}}. Recent empirical studies have also recorded a significant rise in the use of such measures %
\cite{longpre_consent_2024} alongside the growth of a commercial market for services that block online crawlers\footnote{\href{https://blog.cloudflare.com/ai-labyrinth/}{AI Labyrinth}, \href{https://anubis.techaro.lol/}{Anubis}, \href{https://zadzmo.org/code/nepenthes/}{Nephentes}}. The problem with such techno-legal measures, however, is that they are often too crude to capture nuances and complexities in attitudes towards different \textit{purposes} of data crawling. Instead, \texttt{robots.txt}, as an example, offers a catch-all solution to block web-crawlers via a conventional name which not all crawlers faithfully provide nor necessarily conventionally tie to a specific purpose for crawling. This implies that researchers, web archives, and cultural heritage institutions who gather online data for the public good (such as for scientific purposes or to document history) may collaterally be impacted. As a result, the growing resistance against AI crawling and use of anti-web crawling measures may contribute to a gradual closure of the open web \footnote{\href{https://www.technologyreview.com/2025/02/11/1111518/ai-crawler-wars-closed-web/}{AI crawler wars threaten to make the web more closed for everyone}} and damage public, non-commercial data collection. Relatedly, studies have shown that large amounts of content belonging to the public domain -- especially in the area of digitized literature -- is currently handled by commercial actors that make exclusive claims to use the resources for AI training purposes, an issue that raises questions about the privatization and monopolization of public resources in the wake of AI ~\cite{arends_demonopolizing_2024}. As free and public data becomes caged-in and restricted -- meaning new and high-quality data might have to be purchased or produced on demand --  there is a societal risk that power and the ability to innovate in the field of AI will further be concentrated in the hands of a few big actors who have the means to do so. 

\subsection*{3. How can we update technical standards, content licenses, and legal frameworks to better reflect the needs, desires, and consent of organizations hosting content belonging to the commons?}

In response to the increased use of restrictive web-crawling measures, it is becoming increasingly evident that new technical standards are needed; standards that give stakeholders the ability to signal consent in more granular detail. Following Elinor Ostrom's call to protect the commons by establishing clearly defined boundaries for desired and undesired interactions, this would involve developing clear guidelines for what counts as fair and responsible data use and supporting methods for opting-out of training datasets, combined with rules-in-use and appropriate sanctions~\footnote{\href{https://www.mozillafoundation.org/en/blog/a-practical-framework-for-applying-ostroms-principles-to-data-commons-governance/}{A Practical Framework for Applying Ostrom’s Principles to Data Commons Governance}}. At the very least, we need standards that distinguish between AI crawlers, search engine crawlers, and web archiving crawlers. 
We also need better solutions on the other end of the AI pipeline for so-called ``unlearning''~\cite{cooper_machine_2024}, or in the broader sense, the ability to remove content, knowledge, and capabilities from models which should not posses them, whether as the result of withdrawal of data usage consent, or any other valid legal reasons including copyright claims, privacy, and safety concerns. While closely linked to rights enshrined under the GDPR and  necessary for compliance, unlearning on its own should not be seen as an effective mechanism for making outputs compliant with desired policy goals. This is underlined by the fact that for models that have already been trained, and released with open-weights, there are no effective paths for redress via unlearning.

Furthermore, we need updated licensing frameworks that can help ensure that AI training preferences are respected. Such licenses could function as contractual tools to allocate rights and responsibilities relating to  data protection and reuse.\footnote{Also see for example \href{https://licensingafricandatasets.com/open-data-license}{Equitable Open Data Licensing}} A problem with Creative Commons licenses, which constitute the standard for open sharing of content on the web, is that they were designed to promote content dissemination, not restrict it, and they do not specify preferences regarding the purpose of data use.\footnote{\href{https://creativecommons.org/2024/08/23/six-insights-on-preference-signals-for-ai-training/}{Six Insights on Preference Signals for AI Training}. Also see the newly launched CC Signals initiative: \href{https://creativecommons.org/2025/06/25/introducing-cc-signals-a-new-social-contract-for-the-age-of-ai/}{Introducing CC Signals - A New Social Contract for the Age of AI}} While there are ongoing technical governance and standardization initiatives on how to extend and develop technical standards,\footnote{\href{https://datatracker.ietf.org/wg/aipref/about/}{IETF AI Preferences Working Group}, \href{https://datatracker.ietf.org/group/aicontrolws/materials/}{IETF AI Control Workshop}, \href{https://www.w3.org/community/robotstxt/}{W3C Robots.txt Working Group}, retrieved May 9th, 2025.} many of the  working groups are in need of more diverse inputs and increased activity.\footnote{\href{https://almanac.article19.org/guess-who.html}{Internet Standards Almanac}, retrieved May 9th, 2025.}

\subsection*{4. What will the effects of an increased presence of synthetic content in open knowledge databases, and archives belonging to the commons be?} 

Recent research has shown an increased presence of AI generated content on platforms such as Wikipedia~\cite{huang_wikipedia_2025}, highlighting the risk that large quantities of AI generated content (also referred to as synthetic data, or in its lowest quality form, ``AI slop'') may flood online spaces\footnote{\href{https://www.theatlantic.com/technology/archive/2023/03/ai-chatgpt-writing-language-models/673318/}{Prepare for the Textpocalypse}, retrieved June 5th, 2025.}. %
Regardless of whether the presence of such content stems from good intentions or malicious attacks, the need for content moderation on platforms hosting content belonging to the commons is rapidly increasing.
A central problem is that synthetic content often looks like convincing, intentional, and high quality, even though it may contain substantial factual errors, sourcing inaccuracies, and systemic biases. Identifying it is therefore time-consuming and especially challenging for small online communities (such as subcultures or minority language editors contributing to online encyclopedias) who already have very limited resources to devote to fact-checking. Creating confusion regarding the synthetic/non-synthetic nature of content in itself could also be problematic and used as a malicious tool to moderate and censor information~\cite{chesney_bobby_citron_danielle_deep_2019}. As a result, the online presence of AI generated content runs the risk of ``polluting'' shared and open online resources, while causing harm. In addition, the wide-ranging presence of synthetic content online may increase the difficulty of making content belonging to the commons identifiable, visible, and discoverable.\footnote{\href{https://blog.freesound.org/?p=2082}{Freesound in the era of generative Artificial Intelligence}, retrieved June 5th, 2025} This issue is not just important for supporters of the commons, but also AI developers themselves, since uncontrolled presence of synthetic content in web-scraped AI training datasets could result in ``model collapse'' or increasingly more repetitive, biased, and less capable AI. 

\subsection*{5. How can we make sure that the infrastructural and environmental costs of providing data used for AI training are adequately accounted for and fairly distributed?} 

As illustrated by the rising costs of the Wikimedia foundation discussed earlier, the digital commons may be free and open to reuse, but this does not mean that there are no costs associated with their production, maintenance, and distribution. Currently, organizations hosting content belonging to the commons are carrying increasingly heavy financial burdens as a result of AI web crawling. Put bluntly, we may even ask to which extent providers of content belonging to the commons are currently subsidizing AI development by bearing such externalities and costs. Furthermore, significant increases in bandwidth costs just to serve AI crawlers directly translate to an increased environmental footprint which means that digital commons organizations are indirectly (and often involuntarily) contributing to the already problematic environmental effects of GenAI. Finally, the increased burden of moderating and quality control contributes also to the costs imposed by GenAI on the digital commons. How could these infrastructure and environmental costs be better accounted for and compensated? Do AI developers have a responsibility to support the environmental commons and the archives, encyclopedias, forums, and databases on which their "intelligent" products are trained? And if so, what could (or should) such support look like? Unlike copyright protected content, the commons are inherently collaborative and not necessarily ``owned'' or controlled by any single entity who could receive such support. This makes retributions a complex issue. 

\section*{Imagining More responsible Practices}

Currently, we observe a tendency among AI developers to treat the commons as a good or resource that can be exploited \emph{ad infinitum} without cost or harm. While many AI developers fall back on claims concerning model ``openness'' as the mechanism by which they return value to communities from which they extract training data, AI models often remain closed to third-party scrutiny and provide insufficient attribution and recognition to content providers~\cite{widder_why_2024}. Developing AI technologies in ways that truly benefit the public and treat common resources with care would constitute a more responsible practice, requiring further commitments to transparency and accountability around negative externalities. For instance, we may consider developing an "AI commons" and public infrastructures for AI development that would prioritize such commitments, thereby offering a departure from the current status quo. The fact that AI developers often refrain from honoring and returning value to the digital commons from which they extract knowledge~\cite{huang_generative_2023} is on the contrary reminiscent of the tragedy of the commons where a limited number of individuals exhaust a common pool of resources to maximize their own utility.

Instead of viewing the digital commons as a resource or ``thing'' to be exploited, however, we emphasize the need to recognize that extensive labor and resources are needed to keep the commons going. The digital commons should not be treated as an object ripe for extraction, but as an expression of ``\textit{commoning}''~\cite{dulong_de_rosnay_digital_2020}; \ie, as a manifestation of the practice of making, maintaining, and protecting shared and open resources. Commoning is what happens when communities collectively organize, care for, and contribute to the common good and by evoking this term -- which connects to broader discussions on the politics of care %
and has long been discussed in the arts, design, and humanities/social sciences. We call upon the ACM international community to recognize how current developments in GenAI might be negatively impacting the people, communities, infrastructures, and knowledge and information resources that constitute the commons. A first step would be to respect both the legal rules and social norms created by the commoning communities. The value of commoning lies in its ways of supporting open, transparent, and accessible knowledge and information, as well as in how it materializes a decentralized form of knowledge and information governance that is based on collaboration and mutual aid. Let us make sure that current developments in GenAI -- which have a potential to greatly enrich creative practices and support the common good -- does not threaten this unique way of creating, maintaining, and circulating knowledge.

\begin{acks}
This article is the result of a collaborative workshop organized by the Joint Research Center of the European Commission and the fruitful participation of researchers, practitioners, advocates and policy makers on the topic of ``Generative AI and the Future of Digital Common.'' We thank all workshop speakers and participants who enriched this paper with their valuable expertise and views and input.
\end{acks}

\printbibliography

\end{document}